\documentclass[onecolumn,usenatbib]{mn2e}

\usepackage{amsmath,amssymb}
\usepackage{bm}
\usepackage{cases}
\usepackage{soul}

\def\be{\begin{equation}}
\def\ee{\end{equation}}
\def\ba{\begin{eqnarray}}
\def\ea{\end{eqnarray}}

\newcommand{\eref}[1]{Eq.~(\ref{#1})}

\usepackage[dvipsnames]{xcolor}              
\usepackage{graphicx}   
\usepackage{comment}
\usepackage{hyperref}
\hypersetup{
    colorlinks=true,
    linkcolor=red,
    citecolor=blue,
}

\newcommand\nat{Nature}
\newcommand\apjl{Astrophys.~J.~Lett.}
\newcommand\apj{Astrophys.~J.}
\newcommand\prd{Phys.~Rev.~D}
\newcommand\jcap{JCAP}
\newcommand\qjras{Royal~Astronomical~Society,~Quarterly~Journal}
\newcommand\mnras{Month.~Not.~R.~Astro.~Soc.}
\newcommand\prl{Phys.~Rev.~Lett.}
\newcommand\aap{A \& A}
\newcommand\physrep{Physics~Reports}
\newcommand\ssr{Space~Science~Reviews}
\newcommand\aapr{The Astronomy and Astrophysics Review}
\begin{document}

\title{
Intergalactic magnetic field spectra from diffuse gamma rays
}

\date{\today}

\author[W. Chen, B. Chowdhury, F. Ferrer, H. Tashiro, T. Vachaspati]{
Wenlei Chen$^{*}$, Borun D. Chowdhury$^\dag$,
Francesc Ferrer$^{*}$, Hiroyuki Tashiro$^{\circ}$, 
\newauthor Tanmay Vachaspati$^{\dag}$\\
$^{\dag}$Physics Department, Arizona State University, Tempe, Arizona 85287, USA.\\
$^{*}$Physics Department and McDonnell Center for the Space Sciences, 
Washington University, St. Louis, MO 63130, USA.\\
$^\circ$Department of Physics and Astrophysics, Nagoya University, Nagoya 464-8602, Japan.
}

\maketitle

\begin{abstract}
Non-vanishing parity-odd correlators of gamma ray arrival directions observed by Fermi-LAT
indicate the existence of a helical intergalactic magnetic field~\citep{Tashiro:2013ita}. We 
successfully test this hypothesis using more stringent cuts of the data, Monte Carlo simulations
with Fermi-LAT time exposure information, separate analyses for the northern and
southern galactic hemispheres, and confirm predictions made in~\cite{Tashiro:2014}. 
With some further technical assumptions, we show how to reconstruct the magnetic helicity 
spectrum from the parity-odd correlators.
\end{abstract}

\section{Introduction}
\label{sec:intro}

The existence of intergalactic magnetic fields has been speculated for nearly half a century 
largely motivated by the observation of micro Gauss fields in galaxies and clusters of 
galaxies~\citep{1992ApJ...387..528K,2001ApJ...547L.111C,2008Natur.454..302B,2010A&A...513A..30B}.
A primordial magnetic field can be a crucial ingredient for the formation 
of stars and galaxies~\citep{Rees:1987} and its properties can open a window to
the early universe in the domain of very high energy particle 
interactions of matter~\citep{Vachaspati:1991nm,Vachaspati:2001nb,Copi:2008he,Chu:2011tx,Long:2013tha}. 
It is not surprising then that substantial theoretical effort has been devoted to discover ways to
generate primordial magnetic fields~(for recent reviews, see \citealt{2011PhR...505....1K, 2012SSRv..166...37W, 2013A&ARv..21...62D}), understand their implications for the present
state of the universe~(e.g.,~\citealt{aharonian, kim}),
and to observe them using an array of
tools~(e.g., see \citealt{2012PhR...517..141Y} and
references therein).

Until relatively recently, various cosmological observables yielded an upper bound on the
magnetic field strength at the nano Gauss level~(e.g. \citealt{2013arXiv1303.5076P}), and blazar observations 
have placed lower bounds at the $\sim 10^{-16}~{\rm G}$ level~\citep{Neronov:1900zz,Tavecchio:2010mk,Dolag:2010ni,Essey:2011,Chen:2014}. An important
development in the last few years is an appreciation of the importance of magnetic
helicity in the generation and evolution of cosmological magnetic fields. This has led
to the exciting possibility to use the CP odd nature of the helicity to discover intergalactic
magnetic fields~\citep{2004PhRvD..69f3006C,2005PhRvD..71j3006K,
Kahniashvili:2005yp,Tashiro:2013bxa,Tashiro:2014}. 
Further, the helicity of the intergalactic magnetic field provides 
an unmistakable handle by which one can study detailed stochastic properties. 

We have implemented these ideas to observe and measure 
the strength of the intergalactic magnetic field and its correlation functions using diffuse
gamma ray data from the Fermi-LAT satellite\footnote{http://fermi.gsfc.nasa.gov}~\citep{Tashiro:2013ita}. 
Our analysis and (updated) results are reviewed in Sec.~\ref{sec:review} where we also test and confirm 
predictions made in~\citet{Tashiro:2013ita}. 

The successful predictions bolster confidence in the intergalactic magnetic field hypothesis.
Yet there are some other tests that are also essential to rule out other more mundane
explanations related to the observational techniques. Foremost among these is that
Fermi-LAT observations are not performed uniformly on the sky; instead different parts
of the sky are sampled with (slightly) different time exposures. We now take the time
exposure map into account in our Monte Carlo simulations which we use to calculate
statistical error bars. This results in larger error bars but the signal found in~\citet{Tashiro:2013ita}
persists with high significance. Another test is that Fermi-LAT provides 
two data sets for the diffuse gamma ray sky, namely the CLEAN and ULTRACLEAN
data sets, the former being recommended for analyses of the diffuse background but the latter 
is the most conservative.
The analysis in~\citet{Tashiro:2013ita} used the CLEAN data set but,
throughout this paper, we use ULTRACLEAN, with essentially no change in results. A third
test is that a cosmological signal should be present over the whole sky, in particular
it should be  seen in both the northern and southern hemispheres. So we analyze the
north and south data sets separately. The results show a much stronger signal in the
north and a weaker signal in the south. At present we do not have a good explanation for the 
difference in the strengths of the signal.

In Sec.~\ref{sec:QRMH} we turn our attention to connecting the observed correlator of gamma
rays described in Sec.~\ref{sec:review} to the correlation function of magnetic fields. The analytical
framework has already been set up in \citet{Tashiro:2014}. We now use that framework to give a 
``proof of principles'' reconstruction of the intergalactic magnetic field helical correlator. 

We conclude in Sec.~\ref{conclusions}.

\section{Parity odd correlators: test of predictions}
\label{sec:review}

Consider the location vectors, ${\bm n}(E)$, of gamma rays of energy $E$ on the galactic 
sky. As motivated in~\citet{Tashiro:2013ita}, we consider the triple-product correlator at energies 
$E_3 > E_2 > E_1$,
\begin{equation}
Q \left( R ; E_1,E_2,E_3 \right)  = \frac{1}{N_1N_2N_3} 
	  \sum_{i=1}^{N_1}\sum_{j=1}^{N_2}\sum_{k=1}^{N_3} 
                    W_R({\bm n}_i (E_1) \cdot {\bm n}_k (E_3))\, 
		    W_R({\bm n}_j (E_2) \cdot {\bm n}_k (E_3))\,
		    {\bm n}_i (E_1) \times  {\bm n}_j (E_2) \cdot {\bm n}_k (E_3),
\label{Qdef}
\end{equation}
where the indices $i$, $j$, and $k$ refer to different photons at energy $E_1$, $E_2$ and $E_3$ respectively. 
The top-hat window function $W_R$ is given by
\begin{numcases}
{W_R(\cos\alpha) =}
1, \nonumber & \qquad for $\alpha \leq R$
\\
0,  & \qquad otherwise.
\end{numcases}
The statistic can also be written as
\begin{equation}
Q(E_1,E_2,E_3,R)  = \frac{1}{N_3} \sum_{k=1}^{N_3} 
                                   {\bm \eta}_1\times {\bm \eta}_2 \cdot {\bm n}_k(E_3)
\label{Qdef2}
\end{equation}
where
\be
{\bm \eta}_a = \frac{1}{N_a} \sum_{i \in D(n_k(E_3),R)} {\bm n}_i (E_a), \ \ \ a=1,2
\ee
and $D(n_k(E_3),R)$ is the ``disk'' or ``patch'' with center at the location of 
${\bm n}_k(E_3)$ and with angular radius $R$. 

To calculate $Q$ from data, we bin the data in the energy ranges (10,20),
(20,30), (30,40), (40,50) and (50,60)~GeV. (The energies $E$ will refer to the lower
end of the bin.) We also realize that the data will be 
contaminated by gamma rays from the Milky Way and from other identified sources~\citep{Ackermann:2014usa}. 
These are avoided by only considering $E_3$ photons at very high galactic latitudes,
$|b| > 80^\circ$,
and by excising a patch of angular radius $1.5^\circ$ centered on sources identified in the First LAT High-Energy Catalog~\citep{TheFermi-LAT:2013xza}.

\begin{figure}
  \begin{center}
 \includegraphics[width=150mm]{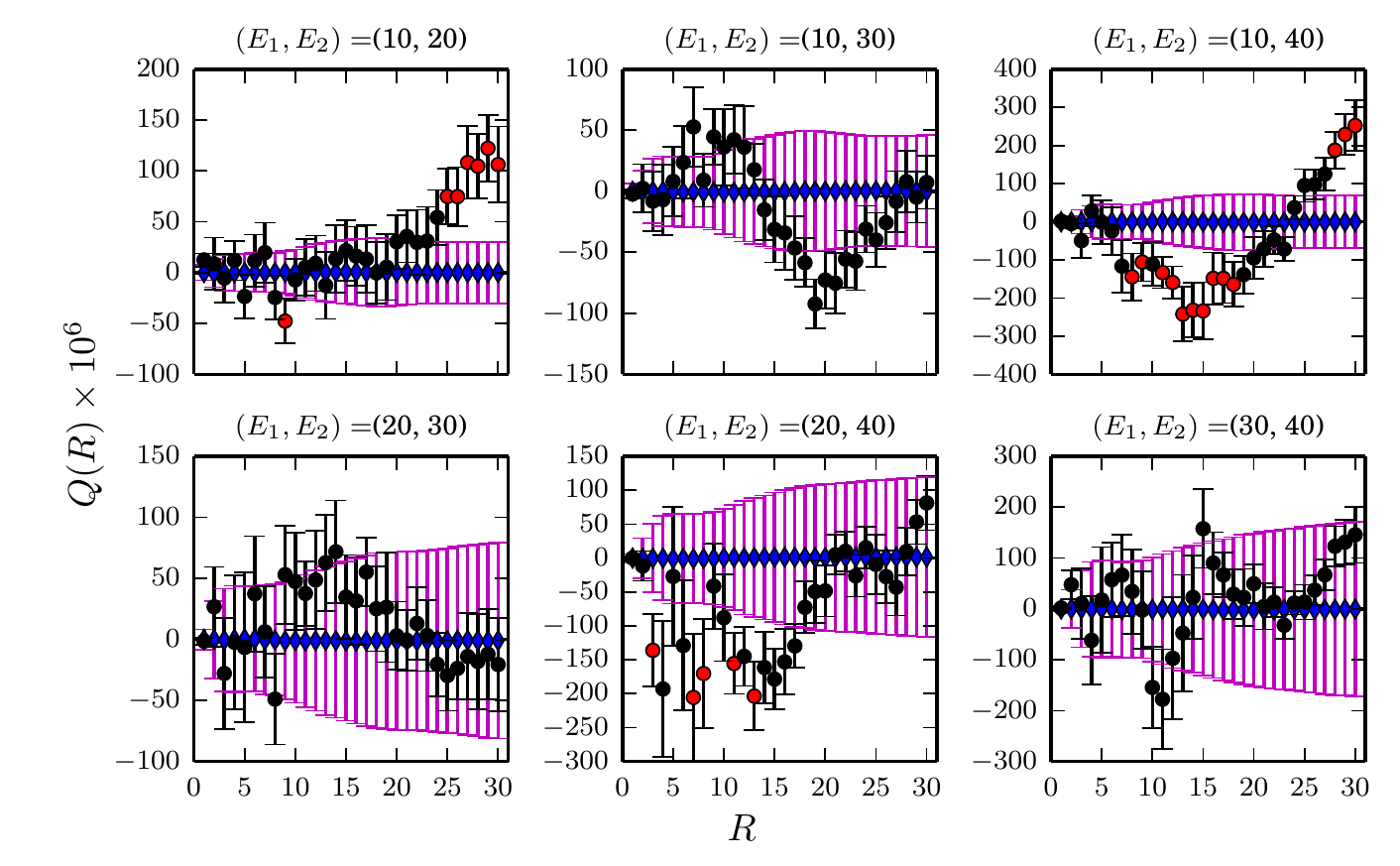}
  \end{center}
\caption{$Q(R)\times 10^6$ versus $R$ for the ULTRACLEAN data set for weeks 9-328 for
$R \leq 30^\circ$. The data points are shown with standard-error error bars. The Monte Carlo %
error bars (magenta) are generated under the isotropic assumption. Data points that deviate 
by more than 2$\sigma$ are colored in red.
}
\label{upto30degrees}
\end{figure}

In contrast to the analysis in~\citet{Tashiro:2013ita} which used the Fermi-LAT CLEAN data set, in this 
paper we will use the more conservative ULTRACLEAN data set for Fermi-LAT observation
weeks 9-328. With the latest release of the Fermi Science Tools, {\tt v9r33p0}, 
we follow the Fermi team recommendations to select good quality data
with a zenith angle cut of $100^\circ$.
However, the results are very similar with the use of either data set. In Fig.~\ref{upto30degrees} 
we show plots
of $Q(R)$ for various energy combinations $(E_1,E_2)$; $E_3$ is always taken to be
$50~{\rm GeV}$. In these plots we also show the spreads obtained from Monte Carlo
simulations assuming that diffuse gamma ray background is isotropically distributed.
In Sec.~\ref{sec:exposure} we will improve on the Monte Carlo simulations by using the Fermi-LAT 
time exposure map.

Another feature in Fig.~\ref{upto30degrees} is that we have extended our analysis to $R=30^\circ$.
This is to test the prediction in~\cite{Tashiro:2014} that $|Q(R)|$ should have a
peak at
\be
R_{\rm peak}(E_2) \approx R_{\rm peak, 0}  \left ( \frac{E_2^{(0)}}{E_2} \right )^{3/2}
\label{Rpeak}
\ee
where $R_{\rm peak, 0}$ is the location of the peak when $E_2 = E_2^{(0)}$.
The plot with $E_1=10~{\rm GeV}$ and $E_2=40~{\rm GeV}$ has a peak at $\approx 14^\circ$.
This implies that all plots with $E_2=40~{\rm GeV}$ should also have peaks at
$\approx 14^\circ$, a feature that is confirmed in Fig.~\ref{upto30degrees}. Then \eref{Rpeak} can
be used with $E_2=30~{\rm GeV}$, $E_2^{(0)}=40~{\rm GeV}$ to predict peaks at $\approx 21^\circ$ 
in the $(10,30)$ and $(20,30)$ plots. This feature is clearly seen in the $(10,30)$ plot in 
Fig.~\ref{upto30degrees}, but a clear peak is not present in the $(20,30)$ plot.
There should also be an inverted peak for $E_2=20~{\rm GeV}$ at $R \approx 40^\circ$. 
However, as we will see in Sec.~\ref{constructnuE},
at such large angles, the Milky Way contribution becomes very strong and we do not see 
any indication of the cosmological signal. We also note that all the identified peaks in 
Fig.~\ref{upto30degrees} are inverted, as we would expect from magnetic fields with 
left-handed helicity provided the spectrum is not too steep~\citep{Tashiro:2014}.

We summarize the peak data in Table~\ref{peaktable} as we will use it in later sections when
we model the magnetic field.

\begin{table*}
  \begin{tabular}{|c||c|c|c|c|c|c| } \hline
   $(E_1,E_2)$ & (10,20) & (10,30) &  (10,40)  & (20,30) & (20,40)  & (30,40) \\ \hline
   $(R_{\rm peak})_{\rm data}$ & ? & $19^\circ$ &  $13^\circ$ & ? &
		       $13^\circ$  &$11^\circ$ \\ \hline
$(Q_{\rm peak})_{\rm data}\times 10^6$ &? &  $-92$ &$-242$  &?  & $-204$& $-177$ \\ \hline
  \end{tabular}
 \caption{
 The peak locations and amplitudes. There is
 no well-identified peak in the (10,20) and (20,30) energy combinations.
 }
 \label{peaktable}
\end{table*}

\section{Time-Exposure and Resampling analyses}
\label{sec:exposure}

\begin{figure}
  \begin{center}
 \includegraphics[width=150mm]{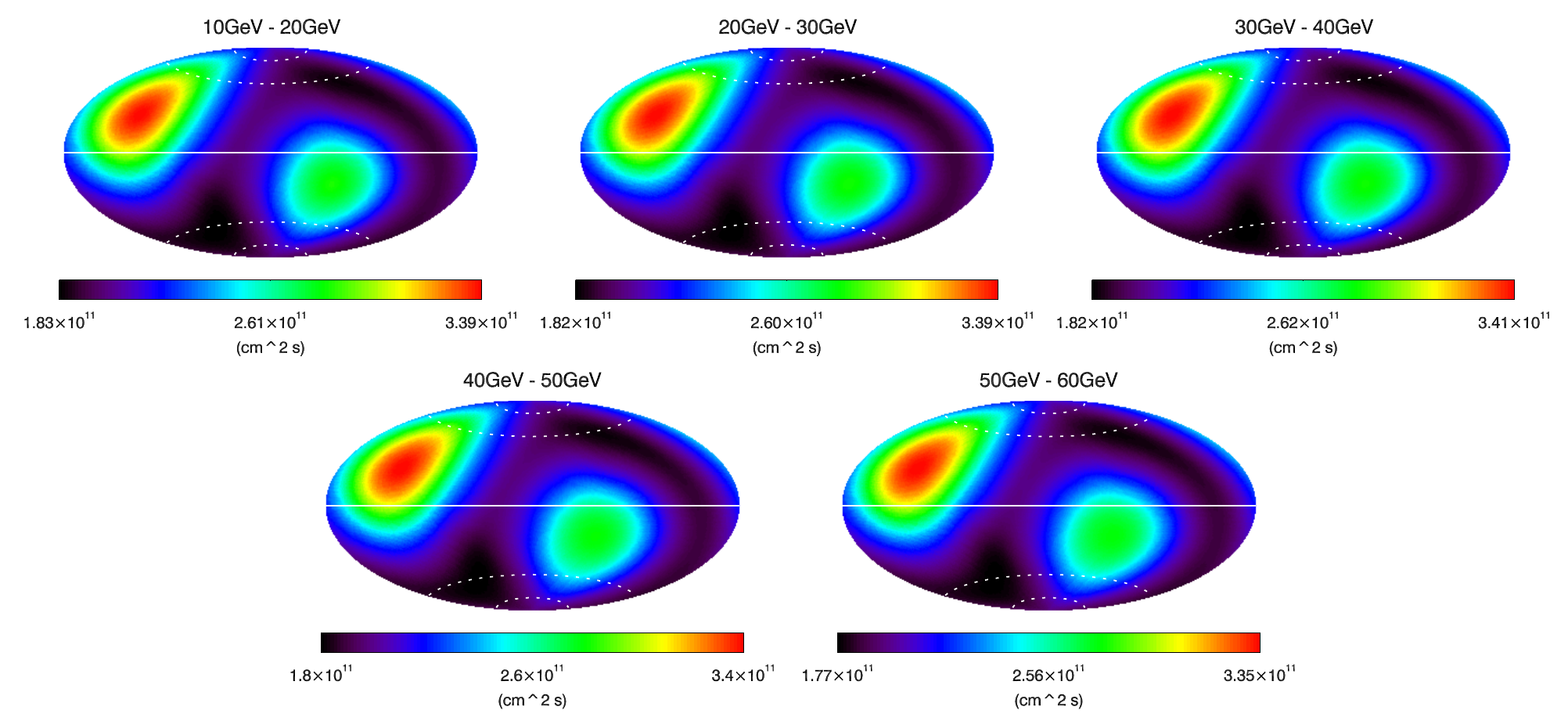}
\end{center}
\caption{Fermi-LAT time exposure maps in the five energy bins. The $|b|=60^\circ, 80^\circ$ galactic
latitudes are shown as dashed white curves. 
 }
\label{timeexposure}
\end{figure}

Fermi-LAT observations do not cover the sky uniformly. Using the latest release
of the Fermi Science Tools, {\tt v9r33p0}, we construct the time exposure map 
by first creating a livetime cube with {\tt gtltcube} and then using
	{\tt gtexpcube2} to obtain full sky exposure maps corresponding to the {\em ultraclean} response function for each 
energy bin.
A plot of the time exposure over weeks 9-328 is shown in 
Fig.~\ref{timeexposure}.

\begin{figure}
  \begin{center}
 \includegraphics[width=150mm]{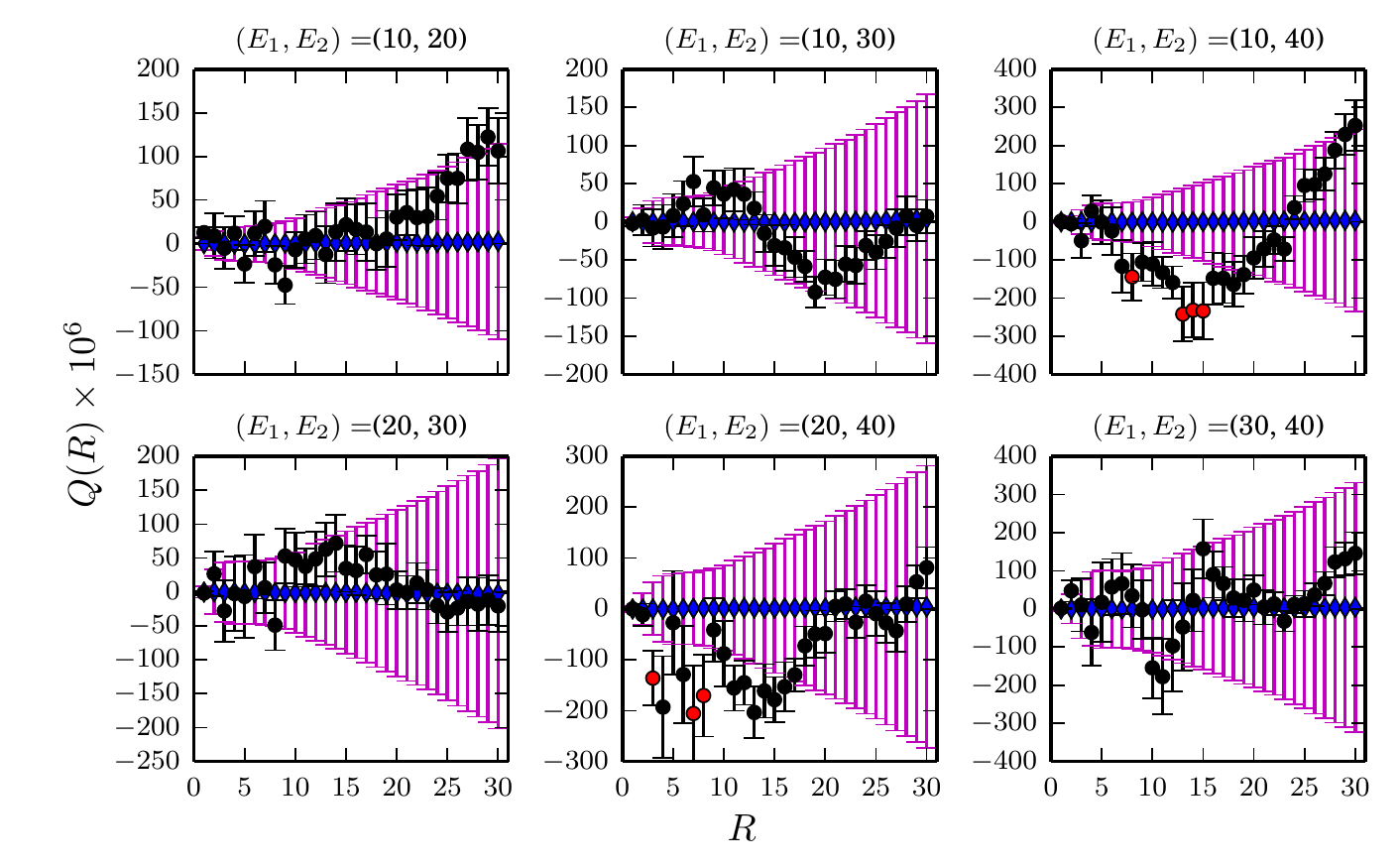}
  \end{center}
\caption{The data overlaid on Monte Carlo generated error bars (magenta) by using the
time-exposure maps shown in Fig.~\ref{timeexposure}.
 }
\label{Qwithexposure}
\end{figure}

With the time exposure maps shown in Fig.~\ref{timeexposure} we run Monte Carlo simulations.
The error bars in Fig.~\ref{Qwithexposure} are the statistical spread in the Monte Carlo $Q(R)$. In 
Fig.~\ref{Qwithexposure} we also overlay the data points on the Monte Carlo error bars. 

\begin{figure}
  \begin{center}
 \includegraphics[width=150mm]{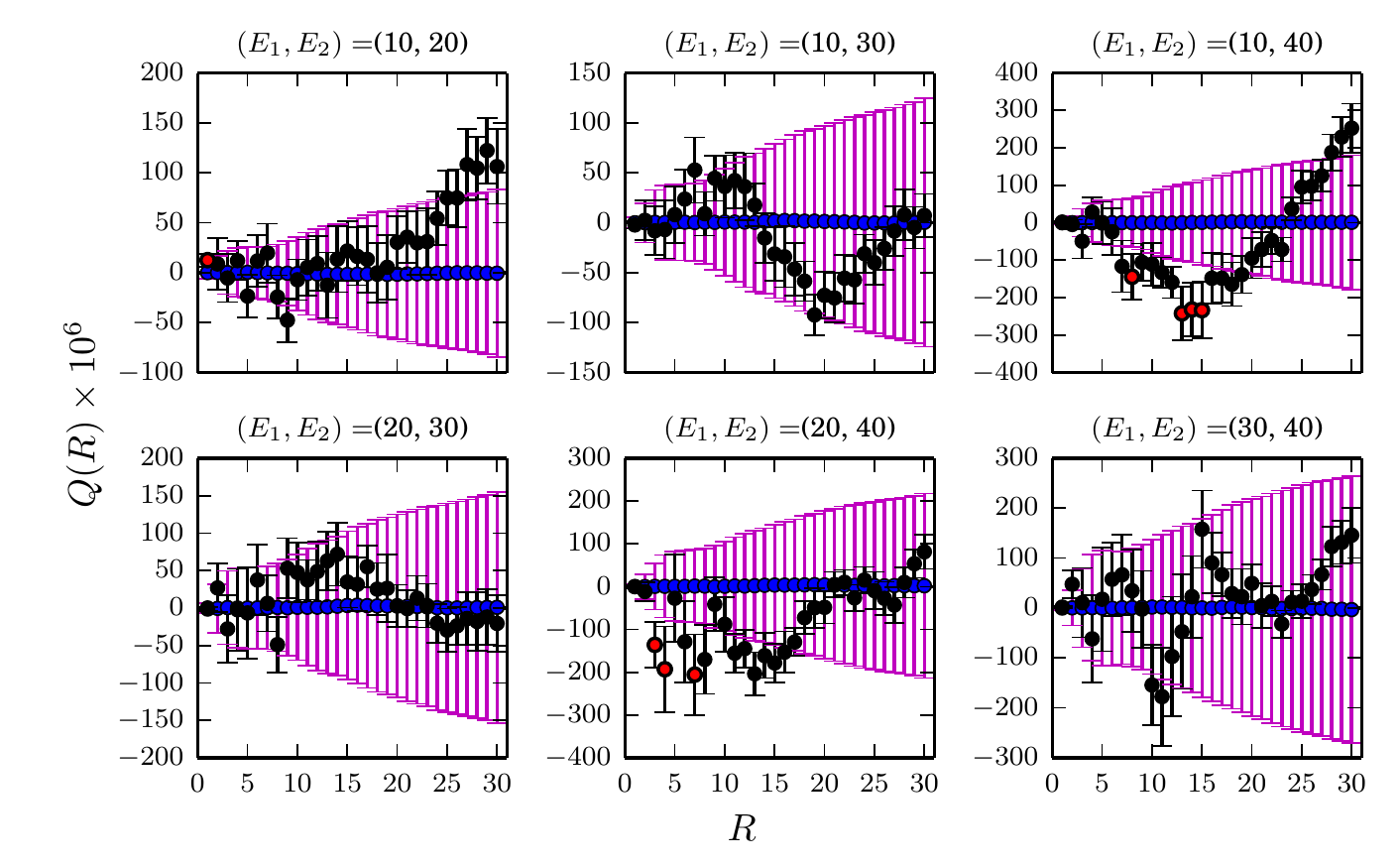}
  \end{center}
\caption{
The data overlaid on Monte Carlo generated error bars (magenta) by using the
resampling method as described in the text.
}
\label{Qwithresampling}
\end{figure}

An alternate way to account for the time exposure is to ``resample'' the data points. 
For this we take the Fermi data and count the number of events in each energy bin. We then strip the data 
of all energy information. Then we randomly resample the observed number counts for each energy
bin from this data set with replacement, choosing the 50~GeV photons first and ensuring that they are 
above $80^\circ$
latitude. $Q(R)$ is then calculated from this resampled data set and the whole procedure is repeated
with 10000 resamples. The result is shown in Fig.~\ref{Qwithresampling} together with the observed
data points. As expected the Monte Carlo error bars are wider than those in Fig.~\ref{upto30degrees} 
and are close to those of the time-exposure results in Fig.~\ref{Qwithexposure}. A disadvantage
in the resampling analysis compared to the time-exposure analysis is that resampling loses any 
energy-dependent factors present in the observations. However, it is an independent check of
the signal.

A cursory look at either Fig.~\ref{Qwithexposure} or Fig.~\ref{Qwithresampling} shows that the 
individual points for $R=13^\circ - 15^\circ$ occurring in the $(10,40)~{\rm GeV}$ energy panel are 
still significant at more than $2\sigma$; while 13 consecutive points, $R=7^\circ-19^\circ$, deviate at 
more than $1\sigma$. However the values of $Q(R)$ are correlated among various values of $R$ and 
also among panels of different energies. In the absence of independent random variables,  the relevant 
quantity to calculate is the probability of the overall patterns of deviations in all the different energy panels.

We have performed a simple evaluation of the statistical significance of the signal by counting
Monte Carlo runs that deviate in the (10,40)~GeV panel by more than $+1\sigma$ at 13 consecutive
$R$ values. This gives a probability of $\simeq 1\%$. We could also include data in the other 
energy panels but have resisted doing so because of the danger that there may be a ``look elsewhere'' 
effect.

While the above test gives an estimate of how likely the data is to occur in the Monte-Carlo runs, 
the criteria relies on the data itself and so is not completely satisfactory. Alternately, as
described in the following sections, our model predicts peaks and also that the peaks are roughly in 
the same locations when $E_2$ is the same as is the case for panels (10,40)~GeV and (20,40)~GeV.  
As another estimate of significance, we sample the Monte Carlo for panels (10,40)~GeV and (20,40)~GeV 
at values of $R=5^\circ,10^\circ,15^\circ,20^\circ$ and ask which of them have peaks at $10^\circ$ or 
$15^\circ$ in both panels. We define a peak by the maximum value being larger than the standard 
deviation at the corresponding value of $R$.  This procedure gives a probability of $\simeq 3\%$.

The correlation between $Q(R)$ for different values of $R$ and also across different panels makes
it difficult to propose a clean test for significance. We have given the result for two such tests. 
However, we expect the issue to become clearer when more data is available and we also plan to 
develop Monte Carlo methods for alternate hypotheses which will allow us to perform Bayesian inference.

\section{North-South analysis}
\label{sec:northsouth}

\begin{figure}
  \begin{center}
 \includegraphics[width=150mm]{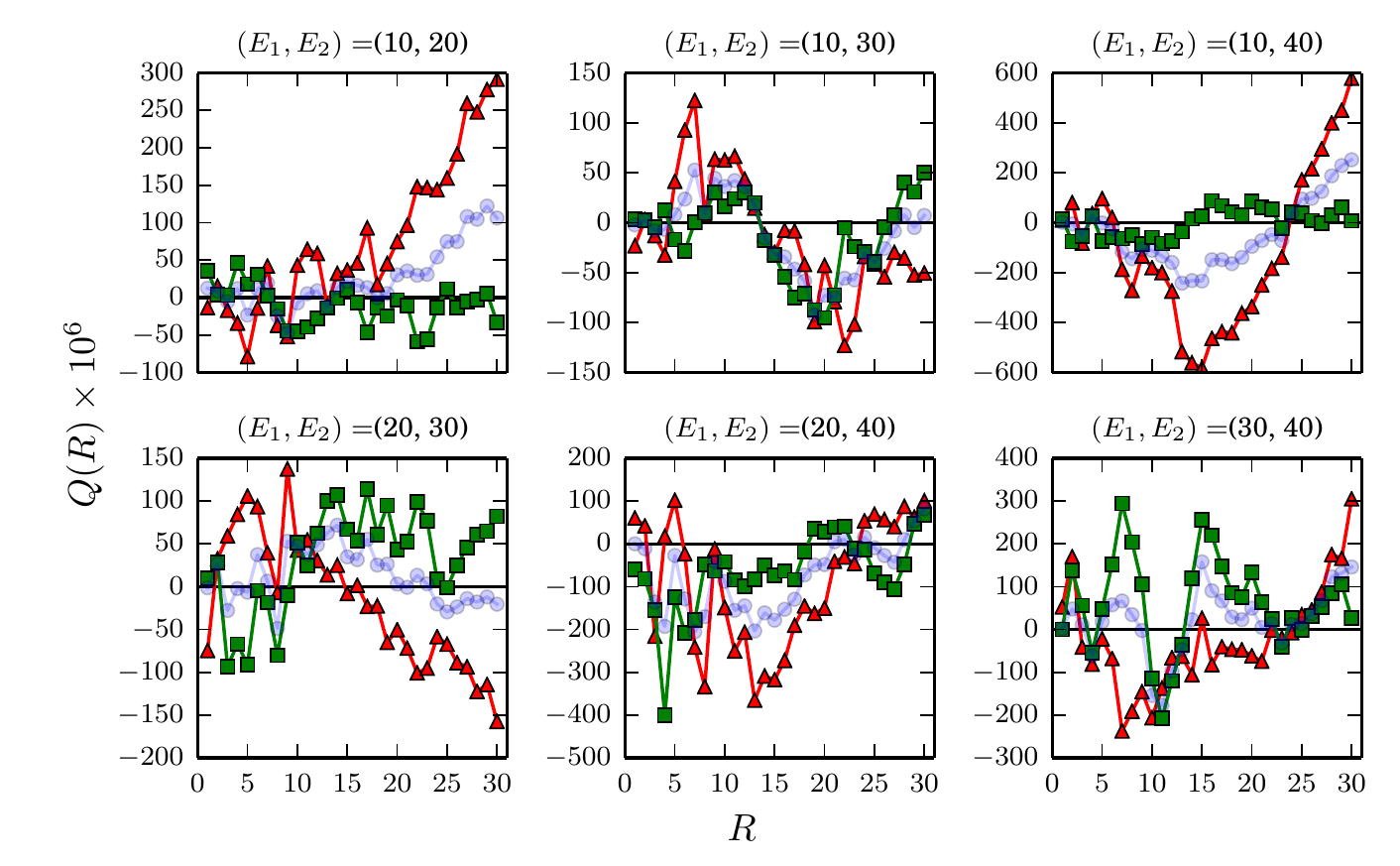}
  \end{center}
\caption{$Q(R)$ vs. $R$ for northern (red), southern (green), and north+south (light blue) data with the
patch center absolute galactic latitudes $> 80^\circ$. We have
not shown error bars for clarity but they are comparable to $\sqrt{2}$ times those shown in 
Fig.~\ref{upto30degrees} with the isotropic assumption and Fig.~\ref{Qwithexposure} with time 
exposure included.
 }
\label{northsouthfig}
\end{figure}

If the signal we see in Fig.~\ref{upto30degrees} is cosmological, we expect  it to exist both
in the northern and the southern hemisphere. So we have split the $E_3=50~{\rm GeV}$
Fermi-LAT data according to whether the galactic latitude is larger than $80^\circ$
(northern hemisphere) or less than $-80^\circ$ (southern hemisphere). The resulting
values of $Q(R)$ are shown in Fig.~\ref{northsouthfig}, where we also plot the full sky results
of Fig.~\ref{upto30degrees} for comparison. 
For clarity, we do not show the Monte Carlo error bars, which will be larger
than those for full sky shown in Fig.~\ref{upto30degrees} by a factor $\approx \sqrt{2}$.

The peak values of $Q(R)$ obtained in the northern polar region have larger amplitude 
than those obtained in the southern polar region. Yet the north-south plots have similarities, 
most striking in the (10,30) and (30,40) panels. 
There can be several reasons for a stronger signal in one polar region than the other.
The existence of the signal depends on the number of TeV blazars in the polar regions. 
The Fermi-LAT time exposure is also slightly different in the north and south, yielding fewer 
photons in almost all bins in the south (see Table~\ref{photonnumber}).

\begin{table*}
  \begin{tabular}{|l||c|c|c|c|c|} \hline
   &10-20 GeV &  20-30 GeV &   30-40 GeV &  40-50 GeV &   50-60 GeV \\ \hline
	  North($>50^\circ$) & 10329& 2598& 1187 & 596 & 350 \\
	  South($>50^\circ$) & 8058 & 2002& 869 & 386 & 267\\ \hline
	  Total ($>50^\circ$) & 18387 &  4600 & 2056 & 982 & 617 \\ \hline \hline
         North($>80^\circ$) &  798 & 208& 119 & 58 & 32 \\
    South($>80^\circ$) & 500 & 124 & 37 & 28 & 23 \\ \hline
    Total ($>80^\circ$) &  1298 &  332 & 156 & 86 & 55 \\ \hline
  \end{tabular}
 \caption{
Number of ULTRACLEAN photons for each energy bin collected in weeks 9-328 and
the North-South distribution for $|b| > 50^\circ$ and $|b| > 80^\circ$ without any source
removal.
}
 \label{photonnumber}
\end{table*}

\section{From $Q(R)$ to $M_H$}
\label{sec:QRMH}

In this section, we will describe how to reconstruct the intergalactic magnetic helicity power spectrum,
$M_H$, from the gamma ray correlators, $Q(R)$. As shown in~\citet{Tashiro:2014}, the plot of $Q(R)$ is
expected to have a peak if there is an intergalactic helical magnetic field. The structure arises
because $Q(R)$ must vanish at small $R$ for mathematical reasons (explained below), and it must also vanish
at large $R$ because non-cascade photons, assumed to be isotropically distributed, dilute the signal 
coming from the cascade photons\footnote{In the realistic case, the Milky Way starts contributing
at very large values of $R$ and can give a non-zero signal. This is the reason why \citet{Tashiro:2013ita}
had restricted attention to $R < 20^\circ$; in Fig.~\ref{upto30degrees} we have gone up to $R=30^\circ$ to
test the prediction of a peak at $\approx 21^\circ$ in the (10,30) and (20,30) panels.}. Therefore,
in between, $Q(R)$ has to have a peak.

In the following analysis, we will adopt the model for $Q(R)$ described in~\citet{Tashiro:2014}.
The first step is to define the bending angle for a gamma ray of observed energy $E_\gamma$~\citep{Tashiro:2013bxa}
\begin{equation}
 \Theta (E_\gamma) \approx
  {e D_{\rm TeV}  D_e \over E_e D_s}  ~{v}_{L} {B}
 \approx
 7.3 \times 10^{-5} ~ \left( { B_0  \over 10^{-16} ~{\rm G}}\right) \left( {E_\gamma
  \over 100 ~{\rm GeV}}\right)^{-3/2} \left( {D_s
  \over 1 ~{\rm Gpc}}\right)^{-1} \left({1+z_s}\right)^{-4},
\label{Theta}
\end{equation}
where $e$ is the electron charge, $v_L \sim 1$ is the speed, and $z_s$ the source
redshift. Second we model the ratio of the number of cascade photons to the total number
of photons in a patch of radius $R$
\be
\frac{N_c}{N_t} = 
\left ( 1 + \frac{0.63~ \nu(E) {\cal A}(R, E)}{1- \exp(-{\cal A}(R,E))} \right )^{-1},
\label{NcbyNt}
\ee
where ${\cal A}(R,E) = A(R)/A(\Theta (E))$ and $A(R) = 2 \pi (1-\cos R)$ is the area of a patch
of radius $R$. \eref{NcbyNt} encapsulates all the information about the noise and signal photons
in the function $\nu(E)$ which can be viewed as the ratio of the number of non-cascade to
cascade photons within a patch of radius $\Theta(E)$. We will estimate this function below.

The model of~\citet{Tashiro:2014} now gives
\be
Q(R) = 
\left ( 1 + \frac{0.63~ \nu_1 {\cal A}(R, E_1)}{1- \exp(-{\cal A}(R,E_1))} \right )^{-1}
\left ( 1 + \frac{0.63~ \nu_2 {\cal A}(R, E_2)}{1- \exp(-{\cal A}(R,E_2))} \right )^{-1}
\frac{1}{1+\nu_3}   (1-e^{-R/\Theta_1})(1-e^{-R/\Theta_2}) Q_\infty ,
\label{QRQc}
\ee
where we are using the short-hand notation: $\nu_a = \nu(E_a)$, $\Theta_a = \Theta (E_a)$
($a=1,2,3$).
The last factor, $Q_\infty$, is related to the magnetic field correlator, 
\begin{align}
Q_\infty =
- \frac{10^{27}}{(1+z_s)^{12} { \rm G^2}}  \left [
 \frac{{q_{12}} |d_{12}| M_H (|d_{12}|)}{{\cal E}_1^{3/2} {\cal E}_2^{3/2}}
+\frac{{q_{23}} |d_{23}| M_H (|d_{23}|)}{{\cal E}_2^{3/2} {\cal E}_3^{3/2}}
-\frac{{q_{13}} |d_{13}| M_H (|d_{13}|)}{{\cal E}_3^{3/2} {\cal E}_1^{3/2}} 
\right ] 
\left ( \frac{1~{\rm Gpc}}{D_s} \right )^2,
\label{Qinfty}
\end{align}
where $q_{12}=\pm 1$ refers to a ``charge ambiguity'' (see Sec.~\ref{reconstructMH}), 
${\cal E}_a = E_a /(10~{\rm GeV})$, and $d_{ab}$ is the distance at
which $Q(R)$ probes the magnetic field
\be
d_{ab} \approx \left ( \frac{\delta_{\rm TeV}}{E_b} \right )^{1/2} - \left ( \frac{\delta_{\rm TeV}}{E_a} \right )^{1/2}
\label{dab}
\ee 
where $\delta_{\rm TeV} \approx 5.6 \times 10^5/(1+z_s)^{4}~{\rm GeV\text{-}Mpc^2} 
\approx 8.8 \times 10^{40}/(1+z_s)^{4}~{\rm Gpc}$.

Given the observation energies $E_1$, $E_2$ and $E_3$, assuming a magnetic field
strength -- more correctly a value for the combination $B_0 (1~{\rm Gpc}/D_s) (1+z_s)^{-4}$
-- we can find $\Theta_1$ and $\Theta_2$
from \eref{Theta}. Note that $\nu_a$ and $Q_\infty$ do not depend on $R$. So we can
(in principle) find the extremum of $Q(R)$ by differentiating \eref{QRQc}. This will tell
us $R_{\rm peak}(E_1,E_2,E_3)$ which we can insert into \eref{QRQc} to get the
ratio $Q_{\rm peak}(E_1,E_2,E_3)/Q_\infty$. If we use data to fix $Q_{\rm peak}$, we can
infer $Q_\infty$ for every energy combination, and that can be related to the magnetic
helicity power spectrum via \eref{Qinfty}. Thus we can reconstruct $M_H$.

A simplified partial analysis of this program yielded the interesting results that $R_{\rm peak}$
is approximately independent of $E_1$ ($E_3$ is fixed for the entire analysis). Hence the
peak position of $Q(R)$ primarily depends on $E_2$ as given in \eref{Rpeak}.
For magnetic power spectra that are not too steep, and if $\nu (E)$ is also constant,
$Q_{\rm peak}$ also only depends on $E_2$~\citep{Tashiro:2014}.

\subsection{Noise to signal ratio}
\label{constructnuE}

The function $\nu (E)$ is a crucial ingredient in the reconstruction of $M_H$. It is defined as
the number of noise photons divided by the number of cascade photons in a patch whose
angular radius is $\Theta (E)$,
\be
\nu (E) \equiv \frac{N_n(\Theta(E))}{N_c(\Theta(E))}.
\label{nuE}
\ee
The noise photons are assumed to be isotropically distributed -- for the time being we ignore
the Milky Way contributions -- and so 
\be
N_n \propto A(R) = 2 \pi (1-\cos R)
\ee
while we expect the cascade photons to be clustered at small $R$. Hence if we plot the total
number of photons in a patch of radius $R$, $N_t(R)$, we expect to see a peak at small $R$, 
and areal growth at large $R$. 

This expectation is, however, confounded by two factors: first, at some large $R$, we expect 
gamma rays from the Milky Way to start out-numbering the cosmological photons, and second,
we have removed Fermi identified sources in our computation of $Q(R)$ to reduce the number 
of non-cascade photons. We will first discuss the Milky Way contamination and find that this
contribution is minimal for $R \lesssim 20^\circ{\rm -}40^\circ$ depending on the energy. The 
source cuts remove both noise and possibly some signal (cascade photons). This fact has
prevented us from deriving $\nu_a$ from the source-cut data. Hence we will leave $\nu_a$
as free parameters  with the additional mild assumption that $\nu_a > 1$, as suggested by
estimates of $\nu_a$ with the data that includes sources.

\begin{figure}
  \begin{center}
 \includegraphics[width=150mm]{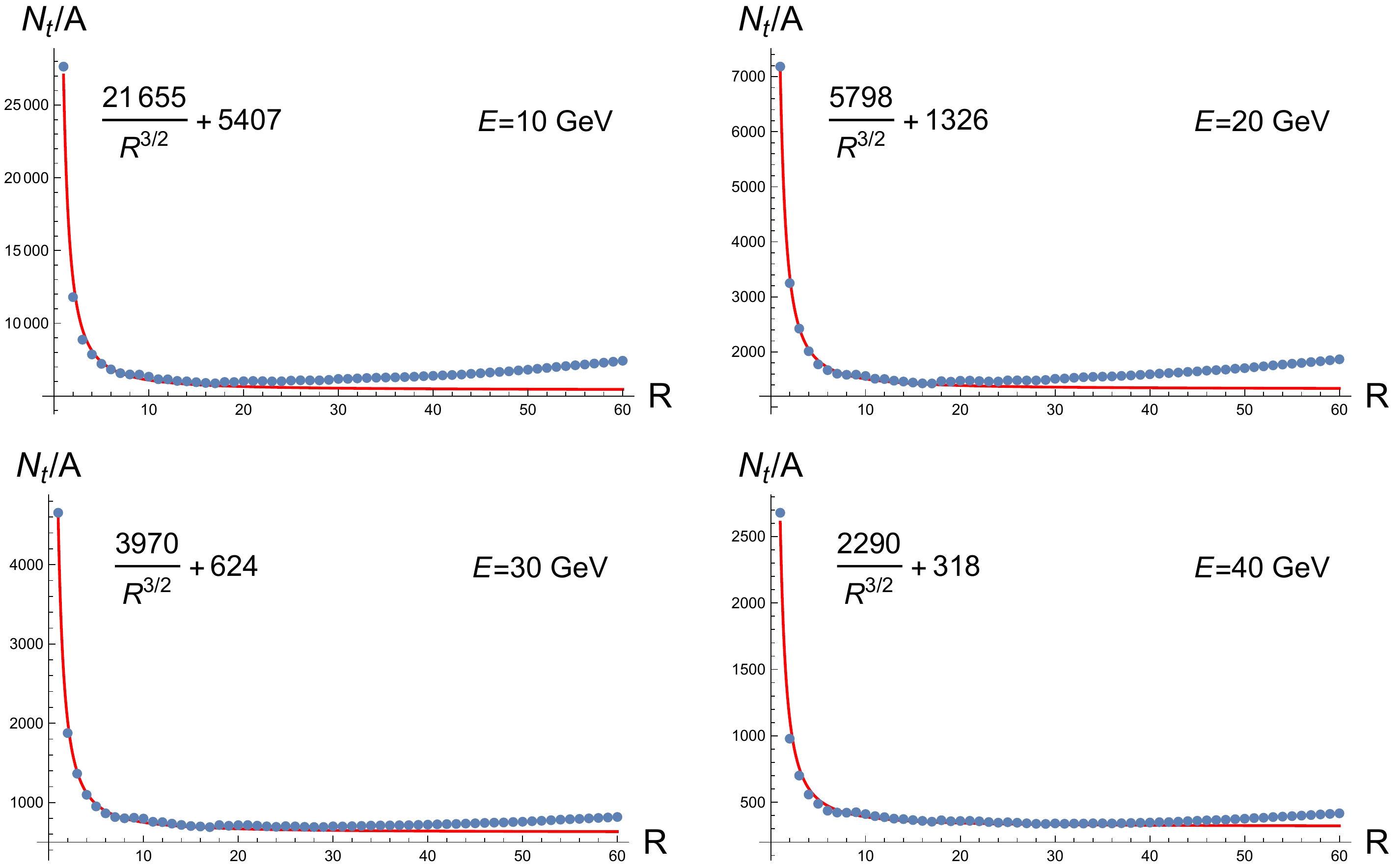}
  \end{center}
  \caption{The average number of photons with energies $10, 20, 30, 40~{\rm GeV}$ in 
high-latitude ($|b|>80^\circ$) patches of size $R$ in the full data {\it i.e.} without source cuts. 
Up to intermediate values of $R$, the data is fit well by the functions shown in the figures and
drawn in red. At larger $R$ the data deviates from the fit because those large patches extend to
lower galactic latitudes where the Milky Way starts contributing to the photon numbers.
 }
\label{NtvsR}
\end{figure}

At some large $R$, we start sampling lower galactic latitudes, and so we expect gamma rays 
from the Milky Way to start out-numbering the cosmological photons. We do not want to extend 
our analysis to such large $R$. To determine these values of $R$, we plot the average number
of photons in a patch of radius $R$ versus $R$, in the data without source cuts. The results
are shown in Fig.~\ref{NtvsR} and matches the clustering at low $R$, a flat
part representing constant areal density, and then a rising part which is due to the Milky
Way contamination. Thus we see that for $E=10,~20~{\rm GeV}$, Milky Way contamination
is minimal for $R \lesssim 20^\circ$, for $E=30~{\rm GeV}$ we have $R \lesssim 30^\circ$,
and for $E=40~{\rm GeV}$ we have $R \lesssim 40^\circ$.
Since we are not interested in gamma rays from the Milky Way, we only consider the curves up 
to the point where they start  rising. Then, as in Fig.~\ref{NtvsR}, we fit the plots to
\be
\left [ \frac{N_t(R)}{A(R)} \right ]_{\rm fit; ~all} = c_1(E) + \frac{c_2(E)}{R^{3/2}}.
\ee
where the subscript ``all'' means that sources are included in these fits.
The coefficient $c_1(E)$ is the areal density of noise photons (including sources), while $c_2(E)$
is related to the number of cascade photons. Thus we get
\be
\nu_{\rm all} (E) = \frac{c_1(E)\Theta(E)^{3/2}}{c_2(E)}
\ee
Note that $\nu_{\rm all}(E)$ depends on the magnetic field through $\Theta(E)$ as given in
\eref{Theta}.

The scheme described above does not
work for $E=50~{\rm GeV}$ photons because our patches are centered on these photons
and there is precisely one $50~{\rm GeV}$ photon per patch. Then to estimate $\nu_{\rm all}(50)$, we
fit a power law to the values of $\nu_{\rm all} (E)$ determined at the lower energies with the
result $\nu_{\rm all} (E) \propto E^{-2.6}$,
and then extrapolate to $E=50~{\rm GeV}$. The resulting numbers for 
$\nu_{\rm all} (E)$ for $B = 5.5 \times 10^{-14}~{\rm G}$
are shown in Table~\ref{nuEtable}. The values for other $B$ are easily determined by
noting that $\nu_{\rm all} (E) \propto \Theta^{3/2} \propto B^{3/2}$ and we can also give an
explicit fitting function
\be
\nu_{\rm fit; all} (E) = 166 \left ( \frac{10~{\rm GeV}}{E} \right )^{2.5}
                             \left ( \frac{B}{5.5 \times 10^{-14}~{\rm G}} \right )^{3/2}
                             \left [ \left (\frac{D_s}{1~\rm Gpc}\right ) (1+z_s)^4  \right ]^{-3/2}.
\label{eq:nuE_50Gev}
\ee

\begin{figure}
  \begin{center}
 \includegraphics[width=150mm]{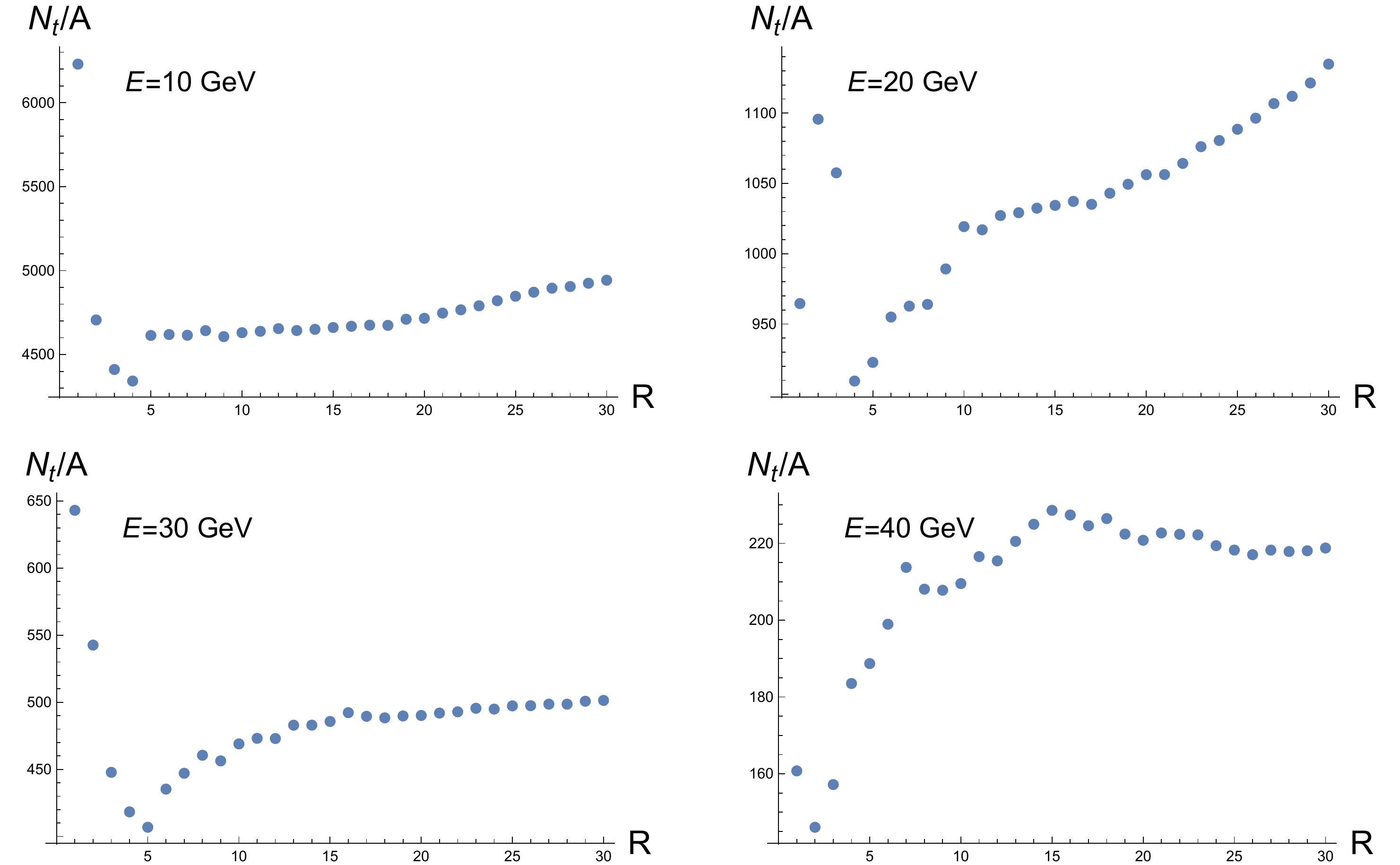}
  \end{center}
\caption{
	Same as in Fig.~\ref{NtvsR}, but now with source cuts.
}
\label{NtnosourcevsR}
\end{figure}

Next we consider the effect of source cuts and plot the average number of photons
in a patch versus patch radius in Fig.~\ref{NtnosourcevsR}. Note that the area of a patch
of radius $R$ is no longer given by $A(R)=2\pi (1-\cos R)$ since the patches have holes
in them. For Fig.~\ref{NtnosourcevsR}, we have used the correct average area of a
patch of radius $R$ that we evaluated by Monte Carlo methods. However, the plots in
Fig.~\ref{NtnosourcevsR} do not have a clear excess at small $R$ that we can identify
with cascade photons, nor a clear flat region that we can identify with a constant areal
density of non-cascade photons. For these reasons, we are unable to extract values
of $\nu_a$ with confidence and will leave these as free parameters for the most part.
Below, when we do need to insert numerical values of $\nu_a$, we will take 
$\nu(E) = \nu_{\rm all}(E)$ as given in Table~\ref{nuEtable}. We hope that the values
of $\nu_a$ will be determined more satisfactorily in the future.

\begin{table*}
  \begin{tabular}{|c||c|c|c|c|c|c| } \hline
   Energy bin & 10-20 &  20-30  & 30-40 & 40-50  &50-60 \\ \hline
$\Theta (E)$ &  $73^\circ$ & $26^\circ$     &$14^\circ$ &$9^\circ$ &
		       $7^\circ$ \\ \hline
$\nu_{\rm all} (E)$ &  156 & 30     & 8    &4 & 2 \\ \hline
\end{tabular}
 \caption{
 Values of the bending angles from \eref{Theta} for $B = 5.5 \times 10^{-14}~{\rm G}$ and 
 $\nu_{\rm all}(E)$ deduced from the fits in Fig.~\ref{NtvsR} that counts ``all'' photons,
 including those from sources. Note that $\nu_{\rm all}(50)$ is 
estimated by using a power law fit to the data for $\nu_{\rm all}(E)$ at other energies.
 }
 \label{nuEtable}
\end{table*}

\subsection{$|{\bf B}|$ from peak position}
\label{Bpeakposition}

To obtain the location of the peaks of $Q(R)$, note that all the $R$ dependent factors in \eref{QRQc} can be
separated out as
\be
q(R) \equiv \frac{(1+\nu_3) Q(R)}{Q_\infty} =
\left ( 1 + \frac{0.63~ \nu_1 {\cal A}(R, E_1)}{1- \exp(-{\cal A}(R,E_1))} \right )^{-1}
\left ( 1 + \frac{0.63~ \nu_2 {\cal A}(R, E_2)}{1- \exp(-{\cal A}(R,E_2))} \right )^{-1}
(1-e^{-R/\Theta_1})(1-e^{-R/\Theta_2}).
\ee
Next we define $x = R/\Theta_1$, $\beta = \Theta_1 /\Theta_2$. We also note that
we are only interested in relatively small $R$ and so $A(R) \approx \pi R^2$. Further,
with the assumption $\nu_i \gtrsim 1$, we can simplify the first two factors around
${\cal A} \approx 1$ and then
\be
q(x) \approx 
\left ( \frac{0.63~ \nu_1 x^2}{1- \exp(-x^2)} \right )^{-1}
\left ( \frac{0.63~ \nu_2 \beta^2 x^2}{1- \exp(-\beta^2 x^2)} \right )^{-1}
(1-e^{-x})(1-e^{-\beta x}).
\ee
The location of the extremum of this function does not depend on the $\nu_a$ since
those are just multiplicative factors. So $q(x)$ will have a peak at $x = x_{\rm peak}(\beta )$.
However, from \eref{Theta}, $\beta = \Theta_1/\Theta_2$ does not depend on $B$.
So the position of the peak in $x$ is independent of the magnetic field, and the peak position
in $R$ depends linearly on $B$ {\it i.e.} $R_{\rm peak} \propto B$. 

\begin{figure}
  \begin{center}
 \includegraphics[width=80mm]{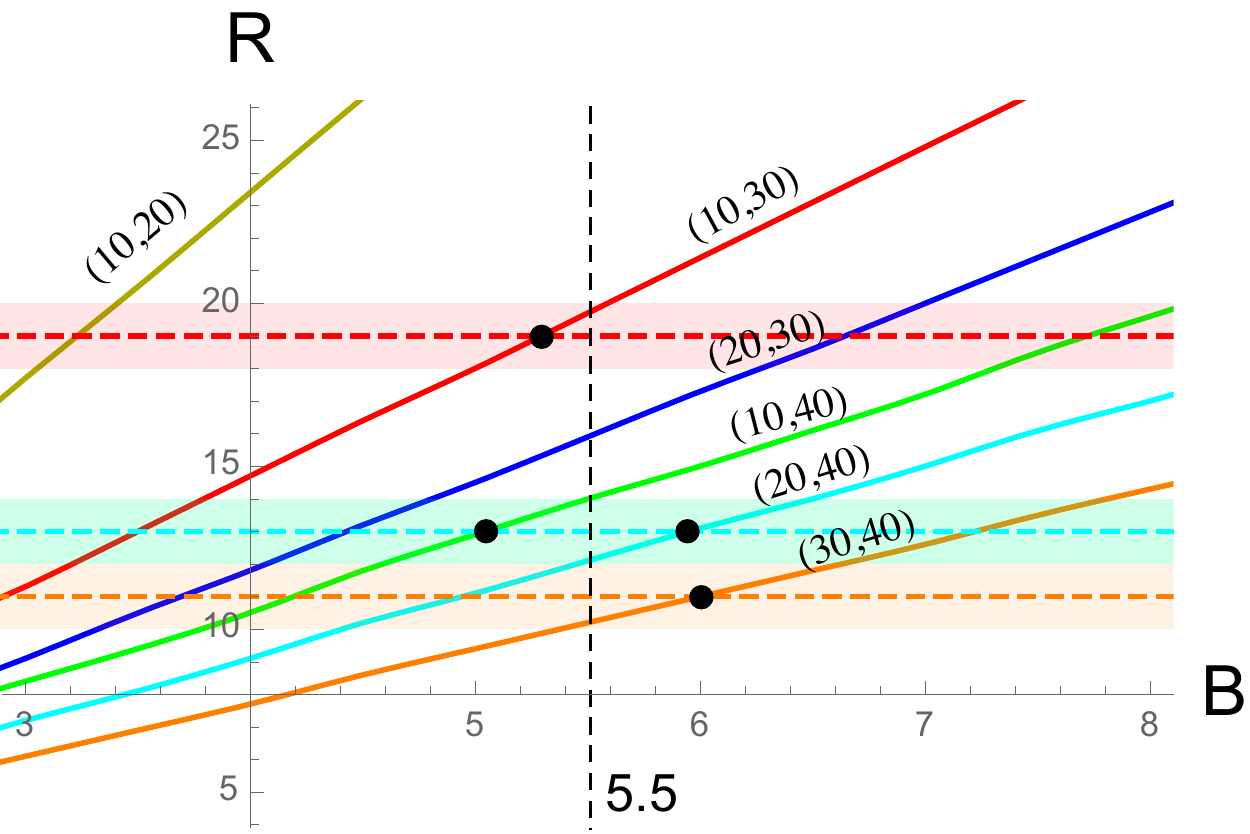}
  \end{center}
\caption{
Peak position versus $B$. The observed values, with an error of $\pm 1^\circ$,
are shown as horizontal bands. The observations are consistent for
$B \approx 5.5\times 10^{-14}({D_s}/{1~\rm Gpc} ) (1+z_s)^4 ~{\rm G}$. 
The plots for $(E_1,E_2) = (10,20), (20,30)~{\rm GeV}$
(top line and blue line)
are drawn though the data does not show any well-defined peaks for these energy
combinations.}
\label{RpeakvsB}
\end{figure}

Fig.~\ref{RpeakvsB} shows the peak position in the model $Q(R)$ as a function of $B$ with the 
choice  $\nu_a =\nu_{\rm all}(E_a)$. Also, the observed  locations of the peaks in Table~\ref{peaktable} 
are shown in Fig.~\ref{RpeakvsB}. We find that all four peak positions line up and can be consistently 
explained with
\be
B \approx 5.5\times 10^{-14} \left [\left (\frac{D_s}{1~\rm Gpc}\right ) (1+z_s)^4  \right ] ~{\rm G} .
\ee
Even though we have adopted $\nu_a =\nu_{\rm all}(E_a)$ for drawing purposes in Fig.~\ref{RpeakvsB}, 
as argued above, the estimate of $B$ is independent of this choice provided $\nu_a \gtrsim 1$.

\subsection{Reconstruction of $M_H$}
\label{reconstructMH}

We now take the values of $\Theta (E)$ from Table~\ref{nuEtable} and the 
peak locations, $(R_{\rm peak})_{\rm data}$, and amplitudes, $(Q_{\rm peak})_{\rm data}$, 
shown in Table~\ref{peaktable}, insert them
in \eref{QRQc} to find $Q_\infty$ for the energy combinations in which a definite peak is
observed. In addition we assume that $\nu_a \gtrsim 1$. Then the values of $Q_\infty$ 
are shown in Table~\ref{Qinftytable}. 
Additionally, we can find the distances $d_{ab}$ for each energy combination
from \eref{dab} with the numerical values shown in Table~\ref{dabtable}.

\begin{table*}
  \begin{tabular}{|c||c|c|c|c|c|c| } \hline
   $(E_1,E_2,E_3=50)$ [GeV] & (10,20) & (10,30) & (10,40) & (20,30)  & (20,40) & (30,40) \\ \hline
$Q_\infty$ &  ? & -2.2 & -4.1 & ? & -0.3 & -0.1 \\ \hline
\end{tabular}
 \caption{
 $Q_\infty$ for different energy combinations with the assumption $\nu(E) = \nu_{\rm all}(E)$,
rounded to one decimal place. 
}
 \label{Qinftytable}
\end{table*}

\begin{table*}
  \begin{tabular}{|c||c|c|c|c|c|c|c|c|c|c| } \hline
   $(E_1,E_2)$ [GeV]& (10,20) & (10,30) & (10,40) & (10,50) & (20,30)  & (20,40) & (20,50)& (30,40)& (30,50)&(40,50) \\ \hline
$d_{ab}$ [Mpc] &  69 & 100  & 118 & 131 & 31 & 49 & 61 & 18 & 31 &12 \\ \hline
\end{tabular}
 \caption{$d_{ab}$ for different energy combinations.
 }
 \label{dabtable}
\end{table*}

The magnetic helicity spectrum is now given by \eref{Qinfty} which we re-write as
\begin{align}
 \frac{m_{ab}}{{\cal E}_a^{3/2} {\cal E}_b^{3/2}}
+\frac{m_{bc}}{{\cal E}_b^{3/2} {\cal E}_c^{3/2}}
-\frac{m_{ac}}{{\cal E}_c^{3/2} {\cal E}_a^{3/2}} 
= -  Q_\infty (E_a,E_b,E_c)
\label{mabeqns}
\end{align}
where ${\cal E}_a = E_a/(10~{\rm GeV})$, $E_a < E_b < E_c $ and
\be
m_{ab} \equiv \frac{ q_{ab} |d_{ab}| M_H( |d_{ab}|) }
     {\left [ 3\times 10^{- 14} (1+z_s)^6  \frac{D_s}{1~{\rm Gpc}} \right ]^2  {\rm G}^2 }
     \label{mabdef}
\ee
In our case, we have observed four peaks, and so we have four such equations,
\ba
\hskip 6 cm
 \frac{m_{13}}{3^{3/2}}
 +\frac{m_{35}}{15^{3/2}}
-\frac{m_{15}}{5^{3/2}} 
&=& 2.2\\
\hskip 6 cm
 \frac{m_{14}}{4^{3/2}}
+\frac{m_{45}}{20^{3/2}}
-\frac{m_{15}}{5^{3/2}} 
&=& 4.1 \\
\hskip 6 cm
 \frac{m_{24}}{8^{3/2}}
+\frac{m_{45}}{20^{3/2}}
-\frac{m_{25}}{10^{3/2}} 
&=& 0.3\\
\hskip 6.2 cm
 \frac{m_{34}}{12^{3/2}}
+\frac{m_{45}}{20^{3/2}}
-\frac{m_{35}}{15^{3/2}} 
&=& 0.1
\label{simultaneous}
\ea
These 4 simultaneous linear equations involve 8 unknowns: 
$m_{13}, ~m_{14}, ~m_{15}, ~m_{24}, ~m_{25}, ~m_{34}, ~m_{35}, ~m_{45}$. 
Thus the system does not have a unique solution and additional physical input is required.
For the present analysis, we will consider three cases: ultra-blue spectrum in which the 
spectrum vanishes at large distances, ultra-red spectrum in which the spectrum
vanishes at small distances, and a power law spectrum.

In the case of the ultra-red spectrum, we assume that the spectrum on the smallest length scales
vanish: $m_{24}=0=m_{34}=m_{35}=m_{45}$ and solve for $m_{13}, ~m_{14}, ~m_{15}, ~m_{25}$.
This does not work, however, because the assumption is inconsistent with \eref{simultaneous},
as the left-hand side vanishes while the right-hand side is non-vanishing. Thus, at least with
these simple assumptions, the data does not fit an ultra-red spectrum.

In the ultra-blue spectrum case, we set
$m_{ab} = 0$ for the larger values of $d_{ab}$ in Table~\ref{dabtable}. Then 
$m_{13}=0=m_{14}=m_{15}=m_{25}$ and the only non-zero unknowns to solve for are
$m_{24}, ~m_{34}, ~m_{35}, ~m_{45}$. The solution is
\be
(d, m)_{45} = (12,367), \ 
(d, m)_{34} = (18,-75), \  
(d, m)_{35} = (31,128),  \ 
(d, m)_{24} = (49,-93), 
 \ \ \ ({\rm ultra-blue\ assumption})
 \label{bluesoln}
\ee
where we also show the distance scale of the correlation (in Mpc).

Next we would like to go from $m_{ab}$ to $|d_{ab}|M_H(|d_{ab}|)$ by using
\eref{mabdef}. Here we need to resolve the discrete ``charge ambiguity'' factors $q_{ab}$.
To understand how these factors arise, recall that pair production by TeV gamma rays
results in electrons and positrons. These carry opposite electric charges and are bent
in opposite ways in a magnetic field. The cascade gamma rays we eventually observe
could have arisen due to inverse Compton scattering of an electron or a positron. This
ambiguity is encapsulated in $q_{ab} = q_a q_b$ where $q_a = \pm 1$ represents the sign of 
the charge of the particle that resulted in the observed gamma ray of energy $E_a$.

To determine $q_{ab}$ we work with the assumption that the magnetic helicity power
spectrum, $M_H(d)$, has the same sign over the distance scales of interest. Then
there are two cases to consider: $M_H > 0$ or $M_H < 0$. If we assume $M_H > 0$,
then with the signs of $m_{ab}$ in \eref{bluesoln},
we get $q_4 q_5 =+1$, $q_3 q_4 =-1$, $q_3 q_5 =+1$ and $q_2 q_4 =-1$. 
Multiplying the first three relations gives $q_3^2 q_4^2 q_5^2 =-1$, which has no real
solutions. Hence there is no consistent solution with $M_H > 0$. On the other hand, if
we assume $M_H < 0$, we need $q_4 q_5 = -1$, $q_3 q_4 =+1$, $q_3 q_5 =-1$ and
$q_2 q_4 =+1$. Then $q_2 = q_3 = q_4 = - q_5$ is a solution (with either choice of 
$q_2=\pm 1$). With this solution, we find
\be
(d_{ab}, |d_{ab}| M_H(|d_{ab}|) =
(12,-3.7), \ 
(18,-0.7), \  
 (31,-1.3),  \ 
(49,-0.9), 
 \ \ \ ({\rm ultra-blue\ assumption})
\ee
where $d_{ab}$ is in Mpc and the correlator is in units of 
$[3\times 10^{-13} (1+z_s)^6 (D_s/1~{\rm Gpc}) ~{\rm G}]^2$. 
However, this estimate of the 
helical power spectrum with the ultra-blue assumption is in tension with the estimate of the 
magnetic field strength
based on the peak location, $\sim 5\times 10^{-14}~{\rm G}$, since the magnetic helicity
density is bounded by the energy density of the magnetic field via the so-called
``realizability condition'' \citep{Moffat:1978}.

Another approach is to assume that the helicity power spectrum has a power law dependence on
distance,
\be
r M_H(r) = A \left ( \frac{r}{\rm Mpc} \right )^p \left [ 3\times 10^{- 14} (1+z_s)^6  \frac{D_s}{1~{\rm Gpc}} \right ]^2  {\rm G}^2 
\ee
and then determine the best fit values of $A$ and $p$ by minimizing the error function
\be
E(A,p) = \sum_{i=1}^4 \left ( \frac{\rm lhs}{\rm rhs} -1 \right)_i^2
\ee
where $i=1,2,3,4$ labels the four simultaneous equations in \eref{simultaneous},
``lhs'' and ``rhs'' refer to the left-hand side and right-hand side of those equations, and we
set $q_{ab}=+1$. Note that the construction of $E(A,p)$ gives equal weight to the four equations;
for example, if we define the error via $\sum ({\rm lhs}-{\rm rhs})_i^2$, this would give greater
weight to equations with larger ${\rm rhs}$. Then we obtain the best fit values
\be
A = 2.08, \ \ \ p = +0.56
\label{MHbestfit}
\ee
which suggests a mildly red spectrum.

We would like to caution the reader that the results of this section depend on the assumption 
$\nu(E) = \nu_{\rm all}(E)$ as given in Table~\ref{nuEtable} which is probably not accurate.
The intent of the above analysis is to show that a reconstruction of the helical power spectrum
may be possible once a definitive model for $Q(R)$ is established.

\section{Conclusions}
\label{conclusions}

We have re-examined the finding in \cite{Tashiro:2013ita} that the parity-odd correlator, $Q(R)$, of 
Fermi-LAT observed gamma rays does not vanish. We have used the most recent data (weeks 9-328)
and find that the signal originally found in data up to September 2013 persists and is, in fact, a
little stronger (see~\url{https://sites.physics.wustl.edu/magneticfields/wiki/index.php/Search_for_CP_violation_in_the_gamma-ray_sky} for a time sequence of results).
 We have also successfully tested the prediction 
of a peak in the (10,30) data around $R \approx 21^\circ$ \citep{Tashiro:2014} and the locations
of the peaks in the other panels as seen in Fig.~\ref{upto30degrees}. Building on the model of
\cite{Tashiro:2014}, we find that all the peak locations in Fig.~\ref{upto30degrees} can be explained
by a single value of the magnetic field strength: $B \sim 5.5\times 10^{-14}~{\rm G}$ 
(see Fig.~\ref{RpeakvsB}). The plots of $Q(R)$ in the northern and southern hemispheres
separately show similar signals in some energy panels but not in others
(Fig.~\ref{northsouthfig}), for which we do not have an explanation.
The peak amplitudes have then been used to reconstruct the
magnetic helicity spectrum under the assumptions of either an ultra-red spectrum or an 
ultra-blue spectrum. The assumption of an ultra-red spectrum does not 
yield a solution; the 
assumption of an ultra-blue spectrum yields the helicity power spectrum at four distance scales,
however the field amplitude is in tension with the realizability condition. 
The sign of the helicity depends on the
sign of the charged particles that are responsible for the inverse Compton scattering of the
CMB photons. We find that this sign ambiguity has a unique resolution if we assume that the
spectrum is either everywhere positive or everywhere negative {\it i.e.} $M_H > 0$ or
$M_H < 0$ at all distance scales, in which case the helicity is  
negative (left-handed). We have also worked with the assumption
that the helicity spectrum is a power law to find the best fit amplitude and exponent as given in
\eref{MHbestfit}. The reconstruction of the power spectrum should be considered a ``proof of principle'' 
and not a definitive claim since it assumes a noise to signal ratio, $\nu(E)$, and also certain other
assumptions, {\it e.g.} that the helical spectrum does not change sign over the length scales of
interest.

We have made several improvements on the
analysis to check the robustness of the results. The change from the Fermi-LAT CLEAN to
ULTRACLEAN data set makes no difference to the signal; the error bars obtained from Monte Carlo
simulations that include the Fermi-LAT time exposure are larger but the signal is still statistically
significant at the $\lesssim 1\%$ level.

Another outcome of our analysis in Sec.~\ref{constructnuE} is that the Milky Way starts 
contributing to the gamma ray data set at $R \gtrsim 20^\circ$ for the 10~GeV bin and at yet 
larger $R$ for higher energies (Fig.~\ref{NtvsR}). Even though our non-trivial signals occur
at $R \lesssim 20^\circ$, this raises the question if the Milky Way is somehow responsible for the 
signal we are detecting. This seems unlikely to us for several reasons:
(i) we observe a signal even in the (30,40)~GeV data set where contamination is seen to be minimal, 
(ii) the signal has a peak structure whereas Milky Way contamination would presumably lead to a
monotonically increasing signal at large $R$, and 
(iii) that the whole pattern of peak locations fits the helical magnetic field hypothesis very well. 
To further confirm the signal, as more data is accumulated, we could restrict attention to only the 
higher energies but reduce the bin size, {\it e.g.} in 5~GeV bin widths instead of the current 
10~GeV bins. The higher energies would limit the Milky Way contamination and the several (smaller) 
bins would still give us the magnetic helicity spectrum over a range of distance scales.

\section*{acknowledgements}
We are grateful to a large number of colleagues who have made suggestions for further tests
of our results in~\citet{Tashiro:2013ita}. We would especially like to thank 
Jim Buckley, Dieter Horns, 
Owen Littlejohns, Andrew Long, Guenter Sigl, David Spergel, and Neal Weiner for their remarks.
This work was supported by MEXT's Program for Leading Graduate Schools ``PhD professional: 
Gateway to Success in Frontier Asia,'' the Japan Society for Promotion of Science (JSPS) Grant-in-Aid 
for Scientific Research (No.~25287057) and the DOE at ASU and at WU.




\end{document}